\documentclass[12pt]{iopart}
\usepackage{iopams}
\usepackage{natbib}
\usepackage{hyperref}
\usepackage{aas_macros}
\usepackage{subcaption}
\usepackage{xcolor}

\newcommand{\rr}[1]{\textcolor{black}{{#1}}}

\newcommand{\rateunits}[0]{\ensuremath{\mathrm{Gpc}^{-3}\,\mathrm{yr}^{-1}}}

\newcommand{\popa}[0]{$\mathcal{A}$}

\newcommand{\popb}[0]{$\mathcal{B}$}

\newcommand{\popc}[0]{$\mathcal{C}$}

\usepackage{graphicx}
\graphicspath{{./Figures/}}

\begin{document}

\title[Cosmic chemistry with XG observatories]{Probing cosmic chemical enrichment with next-generation gravitational-wave observatories}

\author{Maya Fishbach}

\address{Canadian Institute for Theoretical Astrophysics, David A. Dunlap Department of Astronomy and Astrophysics, and Department of Physics, 60 St George St, University of Toronto, Toronto, ON M5S 3H8, Canada}
\ead{fishbach@cita.utoronto.ca}

\begin{abstract}
By observing binary black hole (BBH) mergers out to the edge of the Universe, next-generation (XG) ground-based gravitational-wave (GW) detectors like Cosmic Explorer and Einstein Telescope will map the BBH merger rate across all of cosmic history. 
This merger rate traces the formation rate of their progenitor stars convolved with a delay time distribution. 
Given theoretically-motivated priors on the delay time distribution, we show how XG observations can measure the BBH progenitor formation rate, probing the star formation rate (SFR) up to $z > 15$. 
However, the progenitor formation rate does not directly give a measurement of the SFR, but rather a combination of the SFR and its metallicity distribution as a function of redshift. 
Fortunately, the metallicity-dependence of BBH formation likely varies as a function of BBH mass and/or formation channel. 
We find that if different BBH subpopulations with distinct metallicity biases can be identified, comparing their rates as a function of redshift yields a simultaneous measurement of the SFR and its metallicity distribution. Given optimistic theoretical priors and one year of observation, this may provide a $\sim10\%$ measurement of the SFR at its peak and a 0.2 dex (0.7 dex) measurement of the median metallicity out to $z = 10$ ($z = 15$) at 90\% credibility, although the uncertainties scale with theoretical uncertainties on BBH delay times and formation efficiencies. 

\end{abstract}

%
%
%
%
%

\section{Introduction}
\label{sec:intro}

The current generation of ground-based gravitational-wave (GW) observatories LIGO, Virgo and KAGRA are surveying the population of merging compact object binaries out to redshifts $z \lesssim 2$~\citep{2015CQGra..32g4001L,2015CQGra..32b4001A,2021PTEP.2021eA101A}. In their first three observing runs, the Advanced LIGO-Virgo-KAGRA (LVK) observatory network detected $\approx 70$ binary black hole (BBH) mergers out to $z \approx 1$~\citep{2019PhRvX...9c1040A,2021PhRvX..11b1053A,2023PhRvX..13d1039A}.
At design A+ sensitivity, the LVK network will observe hundreds to thousands of merging BBHs out to $z \approx 2$~\citep{2022arXiv220211048B}.

The growing GW catalog is expanding our understanding of the BBH population and their progenitor stars. 
With hundreds of events, we can map the population properties of BBH systems, including how their merger rate varies with mass, spin and merger redshift, over the Universe's past 10 billion years.
The existing observations are revealing features in the BBH population, including peaks and dips in the mass distribution, correlations between BBH masses and spins, and evolution of the BBH merger rate with redshift~\citep[e.g.][]{2017ApJ...851L..25F,2018ApJ...856..173T,2018ApJ...863L..41F,2021ApJ...913L...7A,2021ApJ...913L..19T,2023ApJ...946...16E,2024PhRvX..14b1005C,2023ApJ...955..107F,2021ApJ...922L...5C,2023PhRvX..13a1048A}. 

The proposed next-generation of ground-based GW observatories, including Einstein Telescope and Cosmic Explorer, would be sensitive enough to detect BBH mergers with total masses above $10\,M_\odot$ out to $z \approx 30$~\citep{2019CQGra..36v5002H,2020JCAP...03..050M,2021arXiv211106990K,2023arXiv230613745E,2023arXiv230710421G,2023JCAP...07..068B}.
As~\citet{2019ApJ...886L...1V} showed, this would enable a direct measurement of the BBH merger rate over the entire history of the Universe, from the very first mergers to the present day. 

The redshift evolution of the BBH merger rate informs a combination of their progenitor formation history and the delay time distribution between progenitor formation and BBH merger.
For the remainder of this work, we assume that BBH systems are the remnants of massive stellar evolution (as opposed to, e.g., primordial BHs;~\citealt{2020ARNPS..70..355C}).
Starting from a massive star origin, the formation of merging BBH systems may involve isolated binary evolution and/or stellar dynamics~\citep[see, e.g.][for reviews]{2021hgwa.bookE..16M,2022PhR...955....1M}.
Isolated binary evolution may include stable mass transfer~\citep{1976IAUS...73...35V,2019MNRAS.490.3740N,2021ApJ...922..110G}, unstable mass transfer leading to a common envelope~\citep{1976IAUS...73...75P,2012ApJ...759...52D}, and chemically homogeneous evolution~\citep{2016MNRAS.460.3545D,2016A&A...588A..50M}.
Meanwhile, dynamical assembly may occur in dense stellar environments such as globular clusters~\citep{1993Natur.364..421K,1993Natur.364..423S,2015PhRvL.115e1101R}, nuclear star clusters~\citep{2016ApJ...831..187A} (possibly involving a central supermassive BH;~\citealt{2022ApJ...929L..22R}), and the disks of active galactic nuclei~\citep{2014MNRAS.441..900M,2023MNRAS.524.2770R}, and involves dynamical friction and three-body interactions leading to binary hardening, exchanges and ejections. 
There are environments in which both binary stellar evolution and dynamical processes are relevant, including stellar triples~\citep{2011ApJ...741...82T,2017ApJ...841...77A} and young star clusters~\citep{2020MNRAS.497.1563R}.  

\subsection{Progenitor formation rate and dependence on metallicity}

Regardless of formation channel, as long as BBH systems have a stellar origin, the BBH progenitor formation rate depends on the properties of stellar populations across cosmic history, including the star formation rate, the stellar initial mass function (IMF), the initial chemical composition (metallicity) of stars, their binary fraction, and the star forming environment, such as whether stars form in a dense cluster or near a supermassive BH.
The conditions of star formation vary significantly over cosmic time~\citep{2014ARA&A..52..415M}. 

The initial metallicity (particularly the iron abundance) of progenitor stars is especially relevant for BBH formation (\citealt{2010ApJ...715L.138B}; see \citealt{2024AnP...53600170C} for a review). 
The high iron opacity in outer stellar layers drives stellar winds, which leads to mass loss and increased radial expansion~\citep{2000ARA&A..38..613K}.
This means that higher-metallicity stars with a greater iron abundance are less likely to retain enough mass in their cores to collapse to a BH, particularly one more massive than $\sim 10\,M_\odot$.
Furthermore, high-metallicity stars are more likely to merge with their stellar companion due to the increased radial expansion, preempting BBH formation in isolated binaries.
Supernova natal kicks, which can disrupt binary systems, may also be larger at higher metallicities. 
The effect of metallicity on BBH formation efficiency is amplified by the other properties of star formation that correlate with metallicity. 
For example, there are indications that the low-metallicity IMF favors more massive stars, which would increase the number of BHs formed~\citep{2023Natur.613..460L,2023ARA&A..61...65K,2024arXiv240407301H}. The fraction of star formation that occurs in dense stellar clusters is also thought to be higher at low metallicity, which increases the rate of BBH mergers driven by stellar dynamics~\citep{2017A&A...606A..85L,2019MNRAS.482.4528E,2019MNRAS.486.5838R,2023MNRAS.525.4456B}.
Thus, within most proposed formation channels, the formation of merging BBH systems is thought to be more efficient at low metallicities compared to high metallicities, although the details depend on the formation channel. 

The theoretical expectation that BBH formation is more efficient at low metallicities is consistent with current observations of BBH mergers at $z \lesssim 1$.
The overall star formation peaks at redshift $z \approx 2$--$3$, while the star-forming metallicity increases monotonically with time (decreasing redshift) as stars enrich their environments with heavy elements during their lives and deaths.
The steep evolution of the BBH merger rate, as inferred from the latest GW catalog GWTC-3~\citep{2023PhRvX..13a1048A}, favors a combination of short delay times and a progenitor formation rate that peaks at higher redshifts than the~\citet{2017ApJ...840...39M} SFR, consistent with the low-metallicity SFR~\citep{2021ApJ...914L..30F,2023arXiv231203316V,2023MNRAS.523.4539K,2023ApJ...957L..31F,2024ApJ...967..142T,2024ApJ...970..128S}.

\subsection{Delay time distributions}
\label{sec:intro-delaytimes}

The delay time for merging binaries refers to the time between the formation of the progenitor stars and the merger of the BBH.
This timescale is dominated by the GW inspiral time, which depends on how closely the two BHs can be brought together by their evolutionary pathway. 
For circular binaries, the inspiral time scales with the initial orbital separation as $\tau_\mathrm{insp} \propto a^4$~\citep{1964PhRv..136.1224P}.
The delay time experienced by a given BBH merger is drawn from a probability distribution referred to as the delay time distribution.
Regardless of formation channel, the delay time distribution is generally predicted to have a long tail towards long delay times because of the steep dependence of the GW inspiral time on the orbital separation: increasing the orbital separation only slightly results in a much longer delay time. 
The specific delay time distribution is determined by the evolutionary pathway.
Within isolated binary evolution, it is generally expected that common envelope can lead to shorter delay times than stable mass transfer because a successful common envelope ejection can shrink the binary orbit more effectively~\citep{2021ApJ...922..110G,2022ApJ...931...17V}.
According to population synthesis models, common envelope evolution generally yields delay time distributions that can be approximated by power laws with slopes $\alpha \approx -1$, while stable mass transfer yields distributions with shallower slopes, although the details depend on uncertain physical parameters~\citep{2023ApJ...957L..31F}.
Within dynamical assembly, the delay time distribution depends on the mass and size of the star cluster, with more massive, denser clusters leading to tighter BBH systems with shorter delay times between star cluster formation and BBH merger~\citep{2018MNRAS.480.5645H,2018ApJ...866L...5R}. 
For a realistic cluster mass function, the predicted delay time distribution is usually well-described by a power-law slope $\approx -1$~\citep{2019PhRvD.100d3027R,2024ApJ...967...62Y}. 

\subsection{Binary black hole subpopulations}
\label{sec:intro-subpop}

In reality, multiple evolutionary pathways, with different progenitor formation rates and delay time distributions, may contribute to the BBH population.
\citet{2021ApJ...913L...5N} argued that a combination of different subpopulations, including the remnants of the first (Pop III) stars, remnants of Pop II isolated binary evolution, and dynamically-assembled binaries in star clusters, may create multiple peaks in the BBH merger rate as a function of redshift.
They showed that these different subpopulations could be identified by measuring the redshift evolution of the BBH merger rate with XG observatories.

In addition to having different merger redshift distributions, BBH subpopulations from distinct formation channels exhibit different mass and spin distributions.
Indeed, there are already indications that the BBH population consists of a superposition of distinct subpopulations from different formation channels, characterized by unique mass, spin and redshift features~\citep{2021ApJ...910..152Z,2021PhRvD.103h3021W,2023arXiv230401288G,2024PhRvL.133e1401L,2024arXiv240403166R}.
BBH systems that were dynamically assembled in dense star clusters, which can be identified by their isotropic distribution of spin orientations~\citep{2016ApJ...832L...2R}, may dominate the merger rate at high component masses $\gtrsim 30\,M_\odot$, perhaps explaining the peak in the BBH mass spectrum at $\sim 35\,M_\odot$~\citep{2023arXiv230401288G,2024arXiv240403166R}. This is in line with some theoretical models~\citep{2023MNRAS.522..466A}.
The highest mass BBH systems may be explained by repeated, hierarchical mergers in dense clusters, which cause these systems to be spinning more rapidly~\citep{2017ApJ...840L..24F,2017PhRvD..95l4046G,2021NatAs...5..749G,2021ApJ...915L..35K,2024PhRvL.133e1401L,2024arXiv240601679P}.
In the future, if orbital eccentricity can be robustly measured from the GW signal, it would provide a powerful discriminator for dynamical assembly that can place constraints on the formation history of globular clusters~\citep{2018PhRvD..98l3005R,2018PhRvD..97j3014S,2021ApJ...921L..43Z,2021MNRAS.506.2362R}.
Among BBH mergers from isolated binary evolution, population synthesis models suggest that evolution involving a common envelope phase may dominate the merger rate at low component masses $\sim 10\,M_\odot$, while stable mass transfer may contribute more at higher masses~\citep{2022ApJ...931...17V}.
Post-common envelope systems may experience tidal spin up, leading to second-born BHs with larger spins~\citep{2018MNRAS.473.4174Z,2020A&A...635A..97B,2021ApJ...921L...2O}.

Even within a fixed BBH evolutionary channel, the formation efficiency and delay time distribution may correlate with BBH mass and spin. 
For example, within isolated binary evolution channels, population synthesis studies often predict that the formation efficiency of high-mass BBH mergers depends most strongly on metallicity, with low-mass BBH mergers (component masses $\lesssim10\,M_\odot$) and BNS mergers exhibiting only mild, if any, dependence on metallicity~\citep[][although see~\citealt{2023ApJ...955..133G}]{2018MNRAS.480.2011G,2018A&A...619A..77K,2019MNRAS.490.3740N,2023MNRAS.524..426I}. 
This means that high-mass BBH mergers probably prefer to form at higher redshifts compared to low-mass systems. 
Meanwhile, the delay time distribution from common envelope evolution may correlate with the spins of the BBH because of the correlation between tidal spin up and the orbital separation of the BBH~\citep{2018A&A...616A..28Q,2022A&A...665A..59B}. 
Dense star clusters probably merge more massive BBHs faster than low-mass BBHs because of mass segregation~\citep{2022ApJ...935..126B,2024ApJ...967...62Y}.

In this work, we make use of the fact that different subpopulations of binary compact object mergers (as identified based on their masses and/or spins) may trace different progenitor metallicities with different delay time distributions. 
We consider a simplified toy model in which the different subpopulations are perfectly identified from their masses and spins.
If we have a theoretical expectation of the delay time distribution corresponding to each subpopulation, we can propagate sources back to their formation redshifts~\citep{2023ApJ...957L..31F}.
Then, by measuring the formation rate of each subpopulation as a function of redshift and comparing them, we can disentangle the overall SFR and its metallicity dependence, as emphasized by~\citet{2024AnP...53600170C}. 

The rest of the paper is structured as follows. In \S\ref{sec:rates-math}, we describe our model for the metallicity-specific SFR and explain its relationship to the BBH merger rate. 
We simulate populations of mock BBH mergers in \S\ref{sec:observations}, discuss their consistency with current LVK data, and show how XG detectors can reconstruct the merger rate as a function of redshift. In \S\ref{sec:inference-SZz}, we describe how to deconvolve the inferred merger rates with a theory-motivated prior on the delay time distribution, yielding a measurement of the BBH progenitor formation rate for each subpopulation (\S\ref{sec:inference-Rf}). Combining this with a theory-motivated prior on the BBH formation efficiency, we show how XG detectors can measure metallicity as a function of redshift (\S\ref{sec:inference-pZ}). We conclude in \S\ref{sec:conclusion}.

\section{From star formation to compact binary coalescence}
\label{sec:rates-math}

\begin{figure}
\caption{\label{fig:SFR-truth} Metallicity-specific SFR using the best fit model reported by~\citet{2023ApJ...948..105V}.}
    \begin{subfigure}[t]{0.5\textwidth}
    \centering
    \includegraphics{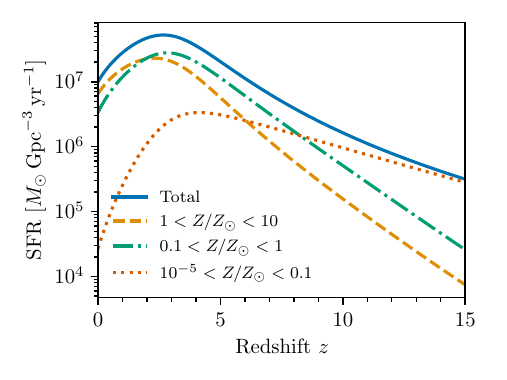}
    \caption{\raggedright Total SFR $\mathcal{R}_\mathrm{SFR}$(blue solid line) assumed in this work, together with the contributions from different metallicity ranges as a function of redshift.}
    \label{fig:SFR_v_redshift_Zbins}
    \end{subfigure}
    \begin{subfigure}[t]{0.5\textwidth}
    \raggedleft
    \includegraphics{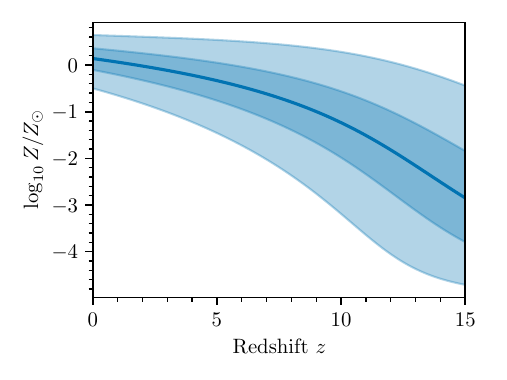}
    \caption{\raggedleft Star-forming metallicity distribution $p(\textcolor{red}{\log_{10} Z} \mid z)$ as a function of redshift. Solid line shows median metallicity at each redshift, while shaded regions contain 50\% and 90\% of the probability.}
    \label{fig:metallicity_versus_redshift_truth}
    \end{subfigure}
\end{figure}

For our mock Universe, we assume that the metallicity-specific SFR, defined as $\mathcal{S}(z, Z) = \mathcal{R}_\mathrm{SFR}(z) p(Z \mid z)$ follows the best-fit model from~\citet{2023ApJ...948..105V}.
In particular,~\citet{2023ApJ...948..105V} fit a~\citet{2014ARA&A..52..415M}-like parameterization for $\mathcal{R}_\mathrm{SFR}(z)$ and a redshift-evolving skewed lognormal distribution for $p(Z \mid z)$ to the Illustris TNG100 simulation~\citep{2018MNRAS.477.1206N,2018MNRAS.475..676S,2018MNRAS.475..648P,2018MNRAS.475..624N,2018MNRAS.480.5113M}. 
Fig.~\ref{fig:SFR_v_redshift_Zbins} shows the the SFR in fixed metallicity bins according to their best fit.
The overall SFR, including all metallicities, peaks at $z \sim 3$. For supersolar metallicities, the peak is at slightly lower redshifts while the low-metallicity SFR peaks at higher redshifts.
Fig.~\ref{fig:metallicity_versus_redshift_truth} shows the metallicity distribution as a function of star-forming redshift $p(\log_{10} Z \mid z)$, \rr{where $Z$ is in units of $Z_\odot$}. The median log-metallicity is represented by the solid line, while the shaded bands enclose 50\% and 90\% of the probability.
Throughout, we assume a solar metallicity $Z_\odot = 0.0142$ from~\citep{2009ARA&A..47..481A} for consistency with~\citet{2023ApJ...948..105V}.
We neglect deviations from the solar chemical abundance as a function of redshift~\citep{2019MNRAS.487.2038C,2022ApJ...925...82K,2022ApJ...925..116S,2024MNRAS.532.3102S}, but we remind the reader that BBH progenitors are most sensitive to iron abundance rather than, e.g., oxygen.

We assume that the BBH metallicity-specific progenitor formation rate $\mathcal{P}(z, Z)$ is related to the SFR by a metallicity-dependent efficiency factor $\eta(Z)$:
\begin{equation}
\mathcal{P}(z, Z) = \mathcal{S}(z, Z)\eta(Z).
\end{equation}
The efficiency, also referred to as the yield, describes the fraction of star-forming mass that ends up in a BBH system that would merge within the age of the Universe, and typically has units $M_\odot^{-1}$ (although see \citealt{2024ApJ...970..128S}, who define a unitless efficiency as the ratio between star-forming mass and BBH total mass). 
The overall progenitor formation rate at a given redshift is $\mathcal{R}_f(z) = \int \mathcal{P}(z, Z)dZ$ (reminiscent of the SFR $\mathcal{R}_\mathrm{SFR}(z) = \int \mathcal{S}(z, Z)dZ$). The BBH progenitor formation rate is a number density (number of BBH progenitors per comoving volume per time), while the SFR is a mass density. 

We write the (in general, metallicity-dependent) delay time distribution as $p_\tau(\tau \mid Z)$.
Recasting all the functions of redshift $z$ in terms of the corresponding age of the Universe at that redshift $t$, the BBH merger rate is:
\begin{equation}
\mathcal{R}_m(t) = \int \int \mathcal{P}(t - \tau,Z) p_\tau(\tau \mid Z) d\tau dZ
\end{equation}
\begin{equation}
\label{eq:Rm_Zdependenttau}
\mathcal{R}_m(t) = \int \eta(Z) \int \mathcal{R}_\mathrm{SFR}(t - \tau)p(Z \mid t-\tau) p_\tau(\tau \mid Z)  d\tau dZ
\end{equation}
If we make the simplifying assumption that $p_\tau(\tau)$ is independent of metallicity $Z$, then the above simplifies to:
\begin{equation}
\label{eq:Rm_givenRfptau}
\mathcal{R}_m(t) = \int \mathcal{R}_f(t - \tau) p_\tau(\tau) d\tau.
\end{equation}
\rr{Although this approximation of a metallicity-independent delay time distribution is unlikely to hold for the full BBH population, we expect it to be a reasonable assumption for individual subpopulations, which is our focus in the following (for further discussion, see \S\ref{sec:inference-Rf} below Eq.~\ref{eq:deconvolution}).}


As discussed in \S\ref{sec:intro}, population synthesis provides predictions for the delay time distribution $p_\tau(\tau \mid Z)$ and the efficiency $\eta(Z)$. 
In the following sections, we explore how to infer the metallicity-specific $\mathcal{S}(z, Z)$ and factorize it into its components $R_\mathrm{SFR}(z)$ and $p(Z \mid z)$ from GW observations of $\mathcal{R}_m$.

\section{Observing the binary black hole merger rate}
\label{sec:observations}

\subsection{Simulated observations}
\label{sec:sims}

We expect that the BBH progenitor formation efficiency $\eta(Z)$ and the delay time distribution $p_\tau(\tau)$ vary as a function of BBH masses and spins. 
For our simulated BBH mergers, we assume a power-law parameterization for $p_\tau(\tau)$ with a minimum delay time $10$ Myr \rr{(corresponding to the typical lifetime of a massive star)} and a slope $\alpha$:
\begin{equation}
    p_\tau(\tau) \propto \tau^\alpha \cdot \Theta(\tau > 10\,\mathrm{Myr}),
\end{equation}
where $\Theta$ is the indicator function.
For $\eta(Z)$, we adopt the parameterization from \citet{2023ApJ...957L..31F}:
\begin{equation}
\eta(Z) = y \cdot \log(w - \log_{10}(Z/Z_\odot)) \cdot \Theta(\log_{10}(Z/Z_\odot) < w - 1).
\end{equation}
Here, $y$ represents the efficiency at low metallicities, which drops off logarithmically towards higher metallicities before cutting off to zero for metallicities above $\log_{10}(Z/Z_\odot)  > w - 1$.
These parameterized models for the delay time distribution and efficiency are approximations to theoretical predictions, but we note that our inference method (introduced in the following \S\ref{sec:inference-SZz}) is insensitive to their precise functional forms, nor does it rely on having a convenient functional form in the first place. 

\begin{figure}
    \caption{\label{fig:dNdz-histogram} Redshift distributions of the mock events from subpopulations \popa{}, \popb{} and \popc{}.}
    \begin{subfigure}[t]{0.5\textwidth}
    \centering
    \includegraphics{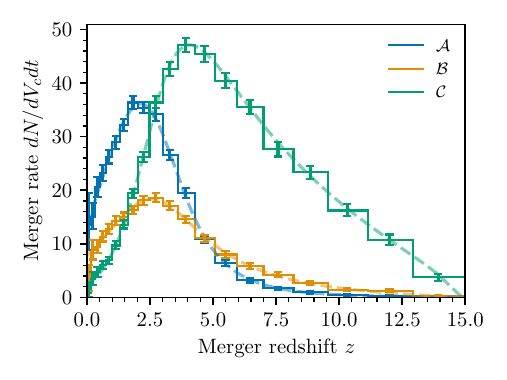}
    \caption{\raggedright Merger rate of each subpopulation as a function of redshift. Dashed lines show the true injected merger rates, while solid piecewise constant lines show the maximum likelihood histogram fit from mock events. Error bars denote 90\% posterior credibility intervals on each bin height.}
    \label{fig:dNdVcdt_3subpops}
    \end{subfigure}
    \begin{subfigure}[t]{0.5\textwidth}
    \centering
    \includegraphics{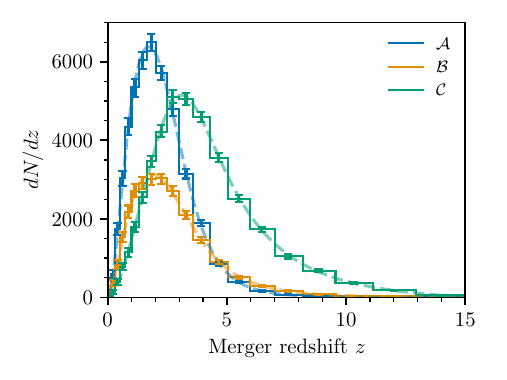}
    \caption{\raggedleft Same as the left panel, but showing the differential number density of mergers as a function of redshift in 1 year of observation.}
    \label{fig:dNdz_3subpops}
    \end{subfigure}
\end{figure}

For simplicity, rather than allowing the efficiency and delay time distribution to vary smoothly across BBH masses and spins, we draw mock BBH mergers from three fixed subpopulations \popa{}, \popb{} and \popc{}, loosely motivated by the following astrophysical formation channels:
\begin{itemize}
    \item Population $\mathcal{A}$: Motivated by low-mass BBH systems produced by isolated binary evolution with a common envelope phase, we assume the formation efficiency is independent of metallicity ($\eta = 10^{-6}\,M_\odot^{-1}$, \rr{chosen to match current measurements of the BBH merger rate described in \S\ref{sec:consistency-lvk}}) and the delay time distribution is a power law with slope $\alpha = -1$. 
    \item Population $\mathcal{B}$: Motivated by more massive BBH systems produced by isolated binary evolution with only stable mass transfer, we assume a shallower delay time distribution with $\alpha = -0.5$ and a metallicity-dependent efficiency with $y = 3\cdot10^{-5}\,M_\odot^{-1}$ and $w - 1 = -0.7$ (so that the efficiency drops to zero above $\approx0.2\,Z_\odot$).
    \item Population $\mathcal{C}$: Motivated by dynamical assembly in dense star clusters, we assume a delay time distribution with $\alpha = -1$ and a stronger metallicity-dependence for the efficiency, with $w - 1 = -1$ (so that the efficiency drops to zero at 0.1 $Z_\odot$) and $y = 8\cdot10^{-5}\,M_\odot^{-1}$. 
\end{itemize}
\rr{The delay time distributions for each subpopulation are chosen to be consistent with theoretical predictions for common envelope, stable mass transfer and dynamical assembly, respectively, as discussed in \S\ref{sec:intro-delaytimes}.}
Note that within each subpopulation, we ignore the metallicity-dependence of the delay time distribution, \rr{applying Eq.~\ref{eq:Rm_givenRfptau} rather than the more general Eq.~\ref{eq:Rm_Zdependenttau}}. We discuss the limitations of this assumption in \S\ref{sec:inference-Rf}. 
Assuming these functions for $\eta$ and $p_{\tau}$ and the metallicity-specific SFR $\mathcal{S}(z, Z)$ described in \S\ref{sec:rates-math}, the merger rate of each subpoulation $\mathcal{R}_m(z)$ is shown as the dashed lines in Fig.~\ref{fig:dNdVcdt_3subpops}.

We now draw events from these three subpopulations.  
Because all BBH mergers with total masses above $\gtrsim15\,M_\odot$ are detectable by XG detectors, we can neglect observational selection effects. 
For our toy model, we only record the redshifts of the mock events, ignoring other properties like masses and spins, and assume that mergers can be perfectly assigned to their subpopulations. In reality, masses and spins will be critical to identify subpopulations. 

We draw merger redshifts from the differential number density $\frac{dN}{dz}$. This number density depends on the merger rate density:
\begin{equation}
\label{eq:dNdz}
\frac{dN}{dz} = R_m(z) \frac{dV_c}{dz} \frac{T_\mathrm{obs}}{1 + z},
\end{equation}
where $dV_c/dz$ is the differential comoving volume, $T_\mathrm{obs}$ is the observing time and the factor of $(1+z)^{-1}$ accounts for time dilation.
The number of events in observing time $T_\mathrm{obs}$ is:
\begin{equation}
N \equiv \int \frac{dN}{dz}dz =  T_\mathrm{obs} \int R_m(z) \frac{dV_c}{dz} \frac{1}{1 + z} dz
\end{equation}

The differential number density for each subpopulation is shown by the dashed lines in Fig.~\ref{fig:dNdz_3subpops}.
We draw 18,000 events from \popa{}, 11,000 events from \popb{} and 24,000 events from \popc{}, corresponding to approximately 1 year of observation.

\subsection{Consistency with LVK observations}
\label{sec:consistency-lvk}

GW observations by the LVK Collaboration have enabled measurements of the BBH population at redshifts $z \lesssim 1$. 
\rr{Taking into account observational selection effects (which, in contrast to XG observatories, are relevant for LVK observations across all masses) and individual events' measurement uncertainties, the BBH merger rate has been measured as a function of BBH mass and redshift.}
The merger rate at $z = 0.2$, the most well-constrained redshift, is around $30\,\rateunits$~\citep{2023PhRvX..13a1048A}. 
Based on measurements of the BBH mass distribution at $z \lesssim 1$~\citep{2023PhRvX..13a1048A}, roughly 60\% of systems in the underlying astrophysical distribution have primary masses around $10\,M_\odot$ (which we identify with \popa{}), 30\% have primary masses around $20\,M_\odot$ (which we identify with \popb{}), and the remaining 10\% have primary masses around $30\,M_\odot$ (which we identify with \popc{}).
Primary masses above $40\,M_\odot$, while common among detected LVK events, make up less than 1\% of the astrophysical population (and therefore the events detected with XG observatories).
The local merger rate derived under our model for each subpopulation (see Fig.~\ref{fig:dNdVcdt_3subpops}) matches these measured rates. 
Furthermore, our association between masses and formation channel is consistent with the recently identified trends between masses and spins and the theoretical predictions discussed in \S\ref{sec:intro-subpop}, although this interpretation remains inconclusive with current data.

In addition to the local rates of the three subpopulations, our model predicts the evolution of the merger rate with redshift, as well as the evolution of the mass distribution with redshift if we believe the mapping between subpopulation and BBH mass described above.
Summing the three subpopulations together, the total merger rate in our model increases with increasing redshift roughly as $(1 + z)^\kappa$ with $\kappa \approx 1$ for $z < 1$. 
This is slightly lower, but still consistent with, the measured redshift evolution from current GW observations. 
In our model, the total merger rate is driven by population \popa{}; population \popc{} has a steeper redshift evolution with $\kappa \approx 2$. 
Assuming \popc{} corresponds to higher BBH masses, this is consistent with existing GW observations; see, e.g.,
Figure 6 in~\citet{2021ApJ...914L..30F} which plots $\kappa$ as a function of BBH primary mass, showing that low-mass BBHs are consistent with $\kappa = 1$ whereas high-mass BBHs may prefer larger values of $\kappa$. Nevertheless, within current observational uncertainties, there is no conclusive evidence that the mass distribution evolves with redshift~\citep{2021ApJ...912...98F,2023PhRvX..13a1048A}. 

\subsection{Inferring the redshift distribution}
\label{sec:merger-rate-inference}

Using the mock events drawn from each subpopulation, we fit the redshift distribution $\frac{dN}{dz}$ with a simple histogram model. Specifically, we model the redshift distribution as piecewise constant in 20 bins equally spaced in $\log(1 + z)$. 
The parameters of the model are the numbers of systems in each redshift bin, or the bin heights $\{n_i\}_{i = 1}^{20}$.
This approach is similar to \citet{2019ApJ...886L...1V}, who used a Gaussian-process regularized histogram~\citep{2014ApJ...795...64F} to fit $\frac{dN}{dz}$ for simulated XG BBH observations.
The Gaussian process regularization places a prior on the bin heights with some correlation length scale between neighboring bins. 
In this work, we take the simplest approach and fit a histogram without any regularization. 
To fit the bin heights $\{n_i\}_{i = 1}^{20}$, we assume that the number of events in each redshift bin $i$ follows a Poisson distribution around the true bin height $n_i$.
We take a Jeffreys prior on each bin height, $p(n_i) \propto 1 / \sqrt{n_i}$. 

For simplicity, we neglect measurement uncertainties for individual events, so that each event can be perfectly assigned to a single redshift bin. 
We do not expect this choice to significantly affect our results. 
Even with realistic measurement uncertainties, we expect most events to contribute to only 1--2 redshift bins (assuming a 3-detector XG network).  
Using the redshift uncertainty prescription reported by~\citet{2019ApJ...886L...1V} for a 3-detector XG network~\citep{2017PhRvD..95f4052V}, the $1\sigma$ measurement uncertainty on $\log z$ is smaller than half a bin width for events below $ z < 4.5$. 
This means that at roughly 68\% credibility, events below $z < 4.5$ (accounting for 94\% of events from $\mathcal{A}$, 85\% of events from $\mathcal{B}$ and 62\% of events from $\mathcal{C}$) can typically be assigned to one redshift bin, events below $z < 9$ (encompassing 99.8\% of events from $\mathcal{A}$, 99\% of events from $\mathcal{B}$ and 95\% of events in $\mathcal{C}$) can typically be assigned to two bins, and the rarest, highest redshift events can be assigned to four bins.  
(These measurement uncertainties do not account for additional uncertainties due to the cosmological model used to convert between GW luminosity distances and redshift, waveform uncertainty, or detector calibration error.) 
Although we neglect them in our toy model, as long as measurement uncertainties are available for each event, they are straightforward to incorporate into the analysis using a hierarchical Bayesian framework, as in~\citet{2019ApJ...886L...1V}.
This may slightly increase our statistical uncertainties on the inferred BBH population, particularly if the individual event uncertainties are larger due to, e.g., fewer detectors in the network, but this can be compensated for by increasing the observing time beyond one year. 

Our fit to the redshift distributions for each subpopulation using the histogram model is shown in Fig.~\ref{fig:dNdz-histogram}.
From our inference of the bin heights $\{n_i\}$, we calculate the differential number density $dN/dz$ by dividing each $n_i$ by the corresponding bin width. 
This is shown in Fig.~\ref{fig:dNdz_3subpops}.
For the differential rate density $dN/dV_cdt$ (Fig.~\ref{fig:dNdVcdt_3subpops}), we divide $dN/dz$ by $(dV_c/dz) T_\mathrm{obs}$ (Eq.~\ref{eq:dNdz}).
The dashed lines show the injected distributions (described in \S\ref{sec:sims}), which are well recovered by the 20-bin histogram models, shown in solid lines. The error bars denote the 90\% Poisson uncertainty on each bin height from the finite number of observed events in that bin.
Our results are comparable to~\citet{2019ApJ...886L...1V}, who show that with 30,000 events detected by XG observatories, the merger rate at its peak redshift can be recovered at the $\approx3\%$ level (68\% credibility), corresponding to $\approx5\%$ at 90\% credibility. We find that at 90\% credibility, the merger rate for populations \popa{} (18,000 events) and \popc{} (24,000 events) can be constrained to $\approx6\%$ at their peak redshifts, while \popb{} (11,000 events) can be constrained to $\approx9\%$, matching the expected $\sqrt{N}$ scaling with number of events. 

\section{Inferring the metallicity-specific star formation rate}
\label{sec:inference-SZz}

\subsection{Propagating mergers back to their formation time}
\label{sec:inference-Rf}

\begin{figure}
    \centering
    \includegraphics{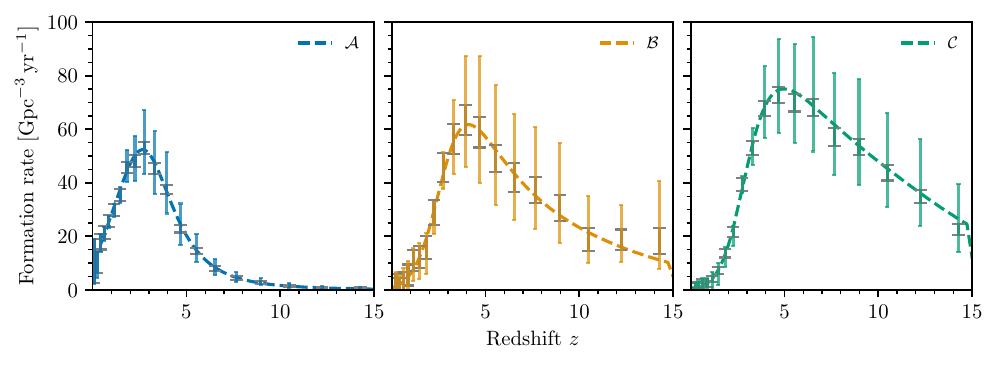}
    \caption{Inferred formation rate for the three different subpopulations. Dashed lines show truth. Error bars show inferred values at 90\% credibility. Gray error bars show the inference when the delay time distribution is perfectly known. Colored error bars show results when marginalizing over a prior on the delay time distribution, \rr{and scale proportionally to the assumed uncertainty on the delay time distribution}.}
    \label{fig:formation_rate}
\end{figure}

If we want to infer the formation rate from the measured merger rate (Fig.~\ref{fig:dNdz-histogram}), we have to deconvolve the delay time distribution, undoing the integral of Eq.~\ref{eq:Rm_givenRfptau}. 
We can numerically approximate the integral over $d\tau$ in Eq.~\ref{eq:Rm_givenRfptau} as a sum over discrete time intervals.
Setting up a grid of times $\{t_i\}$ between zero and the age of the Universe today, we replace the integral with a linear system of equations for each $R_m(t_i)$ in terms of $R_f(t_j)$ and $p_\tau(t_k)$. 
We can then solve this system of equations iteratively for each $R_f(t_i)$ to get the formation rate on the time grid:
\begin{equation}
\label{eq:deconvolution}
    R_f(t_i) = \frac{1}{P_\tau(t_0)} (R_m(t_{i + 1}) - \sum_{j = 0}^{i} R_f(t_j) P_\tau(t_{i + 1 - j})) ,
\end{equation}
where we have defined the delay time probability in the $i$th time interval as $P_\tau(t_i) \equiv (t_{i+1} - t_{i})p_\tau(t_i)$.

Note that this procedure does not depend on any particular parameterization of the merger rate or delay time distribution, as long as each can be evaluated on $\{t_i\}$.
However, this method relies on the approximation that the delay time distribution is independent of formation redshift (in particular, it is independent of metallicity, \rr{as we assumed in going from Eq.~\ref{eq:Rm_Zdependenttau} to Eq.~\ref{eq:Rm_givenRfptau}}).\footnote{We define the delay time distribution to have the same maximum delay time at every formation redshift, even if not all systems that are formed will have time to merge by the present day. Otherwise, if conditioning on systems that merge by $z = 0$, the delay time distribution would always include longer delay times at higher formation redshifts.}
We expect this to be a decent approximation within specific subpopulations, because the relevant physical processes that depend on metallicity will affect $\eta(Z)$ more strongly (see, e.g., Fig. 8 of \citealt{2023ApJ...957L..31F}). In other words, if metallicity strongly affects the evolutionary histories of BBH systems in a specific formation channel, the dominant effect will be to restrict the formation of merging BBH systems to a narrow range of metallicities, over which the delay time distribution will vary less. 
Of course, the total BBH population, encompassing all masses, spins and formation channels, may exhibit a more significant trend between delay times and metallicity. 
If, on the other hand, the delay time distribution varies significantly with redshift, the relation between the merger rate and the formation rate is no longer one-to-one, and one must simultaneously fit the formation rate and the metallicity distribution $p(Z \mid z)$ instead of using Eq.~\ref{eq:deconvolution}. 

We apply the deconvolution of Eq.~\ref{eq:deconvolution} to each subpopulation in order to infer its progenitor formation rate.
We repeat this process for two assumptions about the delay time distribution. First, we assume that the delay time distribution is known perfectly and set it equal to the injected value.
The resulting inference on the formation rate for each subpopulation is shown as the gray error bars (denoting 90\% credibility) in Fig.~\ref{fig:formation_rate}, with the dashed curves showing the true progenitor formation rate.
The locations of the error bars on the redshift axis correspond to the centers of the redshift bins in our histogram fit to the merger rate (Fig.~\ref{fig:dNdz-histogram}).
In this case of perfectly known delay time distributions, the only uncertainty comes from the measurement uncertainty on the merger rate, \rr{and is inherited from the uncertainties in Fig.~\ref{fig:dNdz-histogram}}.
Next, we marginalize over an informed prior on the delay time distribution, assuming a Gaussian uncertainty on the power law slope centered on the injected value with a standard deviation of 0.1. 
(In reality, one could take a set of population synthesis predictions as prior draws, rather than assuming a power-law form for the delay time distribution.)
The corresponding inferred formation rate is overplotted in Fig.~\ref{fig:formation_rate} as the colored error bars ( denoting 90\% credibility). 
\rr{The uncertainty in the delay time distribution propagates to the inferred formation rate, significantly inflating the colored error bars relative to the gray error bars.}
For example, at 90\% credibility, with the 18,000 events from population \popa{}, the progenitor formation rate is measured to within 10\% at redshifts $2 < z < 4$ assuming a perfectly known delay time distribution, or between $24\%$ ($z = 2$) and 43\% ($z = 4$) when marginalizing over our assumed prior on the delay time distribution.
Of course, increasing the prior uncertainty on the delay time distribution would lead to increased uncertainties on the progenitor formation rate inference.

\subsection{Chemical enrichment history}
\label{sec:inference-pZ}

\begin{figure}
    \centering
    \includegraphics{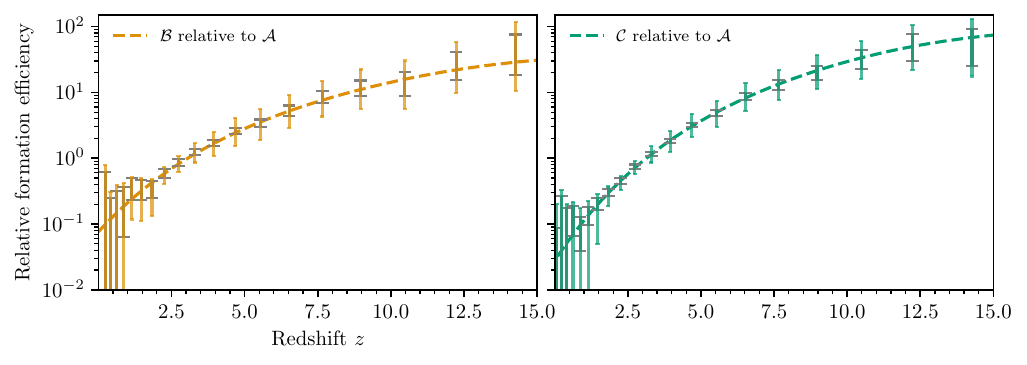}
    \caption{Progenitor formation rates of population \popb{} (left panel) and \popc{} (right panel) divided by that of \popa{} as a function of redshift. Dashed curves show the true relative efficiencies according to our simulation, while the error bars show the results with the mock observations. Gray bars correspond to a perfectly known delay time distribution, while colored bars marginalize over an uncertain delay time distribution.}
    \label{fig:relative_efficiency}
\end{figure}

Recall that the progenitor formation rate depends on the metallicity-specific SFR $\mathcal{S}(z, Z) = \mathcal{R}_\mathrm{SFR}(z)p(Z \mid z)$ and the metallicity-dependent efficiency of the given subpopulation $\mathcal{X}$, $\eta_\mathcal{X}(Z)$ (\S\ref{sec:rates-math}). 
We can disentangle the role of metallicity from the overall SFR by considering the relative formation efficiency between any two subpopulations $\mathcal{X}$ and $\mathcal{Y}$:
\begin{equation}
    \label{eq:relative_efficiency}
    \frac{\mathcal{R}_f^\mathcal{X}(z)}{\mathcal{R}_f^\mathcal{Y}(z)} = \frac{ \int \mathcal{R}_\mathrm{SFR}(z) \eta_\mathcal{X}(Z) p(Z \mid z) dZ} { \int \mathcal{R}_\mathrm{SFR}(z) \eta_\mathcal{Y}(Z) p(Z \mid z) dZ}   
    = \frac{\int \eta_\mathcal{X}(Z) p(Z \mid z) dZ}{\int \eta_\mathcal{Y}(Z) p(Z \mid z) dZ}   
\end{equation}
Note that by dividing the progenitor formation rate of population $\mathcal{X}$ by that of $\mathcal{Y}$, the common factor of $\mathcal{R}_\mathrm{SFR}(z)$ cancels.
We are then left with the cosmic metallicity evolution term $p(Z \mid z)$, which is the same for all subpopulations, and each population's metallicity-dependent efficiency. 

We take population \popa{} as a reference and divide the inferred progenitor formation rates of populations \popb{} and \popc{} by that of \popa{}.
The results are shown in Fig.~\ref{fig:relative_efficiency}.
As in Fig.~\ref{fig:formation_rate}, the gray error bars show the results under a perfectly known delay time distribution, while the colored, larger error bars marginalize over an uncertain delay time distribution. 
The dashed curves show the true, injected relations.
At low redshifts $z \lesssim 1$, the formation rate of all three subpopulations is uncertain because the rate is low (especially compared to the small size of the redshift bins) and not all systems have had time to merge.
At higher redshifts, it is clear that the relative formation efficiencies of both populations \popb{} and \popc{} increase with increasing redshift relative to \popa{}. 
This is because in our example, the progenitor formation of populations \popb{} and \popc{} is more efficient at low metallicity, which is more common at higher redshifts, while the population \popa{} efficiency does not vary with metallicity.

\begin{figure}
    \centering    \includegraphics{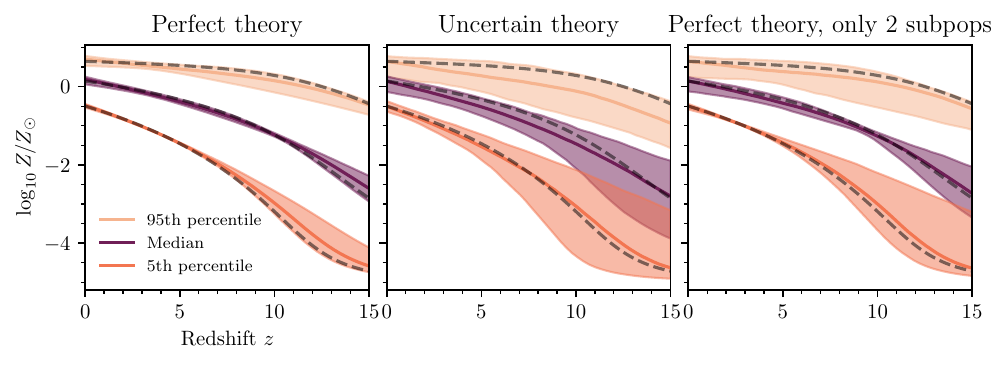}
    \caption{Inferred metallicity distribution as a function of redshift from our mock observations. In each panel, we show the inferred 5th percentile, median, and 95th percentile of the metallicity distribution, with solid lines denoting the median on each quantity and shaded bands showing 90\% credible intervals. Dashed gray lines show the true percentiles, equivalent to what is shown in Fig.~\ref{fig:metallicity_versus_redshift_truth}. Left panel assumes the delay time distribution and metallicity-dependent efficiencies are known perfectly for all three GW subpopulations. Center panel marginalizes over a prior on the delay time distributions and metallicity-dependent efficiencies of the three populations. Right panel assumes perfect knowledge of the delay time distributions and metallicity-dependent efficiencies, but that there are only two subpopulations, $\mathcal{A}$ and $\mathcal{B}$. }
    \label{fig:inferred_metallicity_v_redshift}
\end{figure}

In order to recover the redshift-dependent metallicity distribution $p(Z \mid z)$ from the relative formation efficiencies of Fig.~\ref{fig:relative_efficiency}, we must make some assumptions about $\eta_\mathcal{X}(Z)$ for $\mathcal{X} \in \{ \mathcal{A}, \mathcal{B}, \mathcal{C} \}$.
Given a parametric model for $p(Z \mid z)$ and a prior on $\eta_\mathcal{X}(Z)$, we can fit the measured relative efficiencies as a function of redshift (the error bars in Fig.~\ref{fig:relative_efficiency}) to Eq.~\ref{eq:relative_efficiency} and recover the parameters of the $p(Z \mid z)$ model.
Because our earlier analysis did not build in any correlations between neighboring redshift bins, this procedure is equivalent to fitting a curve parameterized by Eq.~\ref{eq:relative_efficiency} to the uncertain measurements represented by the error bars in Fig.~\ref{fig:relative_efficiency}. 
Indeed, in the following, we approximate these error bars as independent Gaussians (although our method can easily generalize to more complicated, correlated measurements).
If we consider only two subpopulations (a single panel of Fig.~\ref{fig:relative_efficiency}), our likelihood is simply a product of 20 Gaussians (where 20 is the number of redshift bins initially chosen).
When we consider all three subpopulations (both panels of Fig.~\ref{fig:relative_efficiency}), our likelihood is a product of 40 Gaussians.

With this approach, we fit for $p(Z \mid z)$ and $\eta_\mathcal{X}(Z)$ as follows.
We assume the parameterization from \citet{2023ApJ...948..105V} for $p(Z \mid z)$, taking flat priors on all parameters between the ranges quoted in their Table 2, except the skewness parameter (their $\alpha$) which, as discussed later, we hold fixed to its best-fit value.
Recall that our simulated observations were generated assuming this metallicity model with the best-fit parameters from \citet{2023ApJ...948..105V}, as shown in~Fig.~\ref{fig:metallicity_versus_redshift_truth}.
For $\eta_\mathcal{X}(Z)$, we consider the optimistic scenario that it is known perfectly for all three subpopulations a priori, as well as a scenario where we place a theory-informed prior on $\eta_\mathrm{X}(Z)$.
For the theory informed prior, we adopt the same parameterization as our simulations in terms of $y$ and $w$. We assume that the relative heights $y_\mathcal{B}/y_\mathcal{A}$ and $y_\mathcal{C}/y_\mathcal{A}$ are known to within 50\% of their true values, so that their priors follow a lognormal distribution with width 0.5. 
We assume that $\eta_\mathcal{A}$ is known to be constant (independent of metallicity), while for $\eta_\mathcal{A}$ and $\eta_\mathcal{B}$, the metallicity cutoff $w - 1$ is uncertain within 0.1 $Z_\odot$, so that the prior is a Gaussian with standard deviation 0.1 $Z_\odot$. 
As one example, these uncertainties can be compared to the $\eta(Z)$ found for different physics variations in the population synthesis study of \citet{2022ApJ...940..184V}, reported also in Fig. 4 of \citet{2023ApJ...957L..31F}.
According to this study, the current theoretical uncertainty on $y$ is a bit smaller (standard deviation on order 10\%) than what we assume here (50\%) while the current uncertainty on $w - 1$ is larger (standard deviation around 0.2 $M_\odot$, compared to the 0.1 $M_\odot$ we assume). 
Similarly to the choice of prior on the delay time distribution, one could easily replace our assumed functional form for $\eta(Z)$ with prior draws directly from population synthesis.   

The corresponding inference on the metallicity distribution as a function of redshift is shown in Fig.~\ref{fig:inferred_metallicity_v_redshift}. 
From the resulting posteriors on the parameters describing $p(Z \mid z)$, we plot the inferred 5th, 50th (i.e. median) and 95th percentiles of the metallicity distribution at each redshift (different colors).
For each metallicity percentile, we show the median (solid line) and 90\% credible intervals (shaded bands).
The true 5th, 50th and 95th percentiles are shown by the dashed gray lines (matching the solid line and outer bands of Fig.~\ref{fig:metallicity_versus_redshift_truth}). 
The leftmost panel corresponds to the case where the delay time distribution $p_\tau^\mathcal{X}$ and efficiency $\eta_\mathcal{X}(Z)$ are both known perfectly, while the middle panel marginalizes over our assumed theory-informed priors on $p_\tau^\mathcal{X}$ and $\eta_\mathcal{X}(Z)$.
Both of these panels use all three subpopulations $\mathcal{X} \in \{\mathcal{A}, \mathcal{B}, \mathcal{C}\}$.
In the rightmost panel, we assume we only have subpopulation \popb{} to compare against \popa{}. 

The true chemical enrichment history is recovered within statistical uncertainties in all three cases.
We note that our priors on the parameters of the $p(Z \mid z)$ distribution, taken from \citet{2023ApJ...948..105V}, place more prior volume for narrower metallicity distributions (smaller $\omega_0$ and $\omega_z$) compared to the injected value, which falls near the upper edge of the prior.
For this reason, the posterior on the 5th and 95th percentiles of the metallicity distribution are also skewed to higher and lower values, respectively, relative to the injected value. 
In the limit that the delay time distribution $p_\tau^\mathcal{X}$ and metallicity-dependent efficiency $\eta_\mathcal{X}(Z)$ are known perfectly for all three subpopulations, we can recover tight constraints on the metallicity distribution $p(Z \mid z)$. 
In this case, corresponding to the left panel of Fig.~\ref{fig:inferred_metallicity_v_redshift}, the median log-metallicity is measured to within 0.66 dex at all redshifts (or to within 0.19 dex for $z < 10$).
(These values, and the following summary statistics, correspond to the width of the corresponding inferred 90\% credible interval.)
Meanwhile, the redshift evolution of the average metallicity, summarized by the difference in median log-metallicity between $z = 15$ and $z = 0$, is measured to within 19\% of its median inferred value. 
The width of the metallicity distribution (i.e. the difference between the 95th and 5th percentiles) is measured to within 36\% across all redshifts.   

However, these predicted constraints depend on how well $p_\tau^\mathcal{X}$ and $\eta_\mathcal{X}(Z)$ are known for each subpopulation. The delay time distribution affects how well we can reconstruct the progenitor formation rates (Fig.~\ref{fig:formation_rate}) and compare them between different subpopulations (Fig.~\ref{fig:relative_efficiency}). 
Meanwhile, the metallicity-dependent formation efficiency affects how well we can translate the relative formation efficiencies as a function of redshift (Fig.~\ref{fig:relative_efficiency}) to the metallicity distribution at each redshift (Fig.~\ref{fig:inferred_metallicity_v_redshift}; see also Fig. 5 in \citealt{2023ApJ...957L..31F} for an illustrative example with current GW data). 
If we assume that $p_\tau^\mathcal{X}$ and $\eta_\mathcal{X}(Z)$ are imperfectly known, corresponding to the second panel in Fig.~\ref{fig:inferred_metallicity_v_redshift}, our uncertainty on the metallicity distribution unsurprisingly increases.
Marginalizing over our assumed priors on $p_\tau^\mathcal{X}$ and $\eta_\mathcal{X}(Z)$, the median log-metallicity is measured to within 2.0 dex (or 1.0 dex for $z < 10$), while the width of the metallicity distribution is measured to within 97\%.  
The difference in median log-metallicity between $z = 15$ and $z = 0$ is measured to within $71\%$.
Note that if we adopted different theoretical uncertainties on $p_\tau$ or $\eta$, the uncertainties on $p(Z \mid z)$ would shrink or expand accordingly. 

If we had two, rather than three, GW subpopulations to compare against each other, we expect worse constraints on the metallicity distribution. 
The rightmost panel of Fig.~\ref{fig:inferred_metallicity_v_redshift} shows this scenario for subpopulations $\mathcal{A}$ and $\mathcal{B}$, assuming their delay time distributions and metallicity-dependent efficiencies are known perfectly.
The median metallicity is measured to within 1.3 dex (0.37) dex at all redshifts ($z < 10$), the width of the metallicity distribution is measured to within 75\%, and the difference in median metallicity between $z = 0$ and $z = 15$ is measured to 48\%. 
The absence of subpopulation $\mathcal{C}$ means that there are significantly fewer high-redshift events, contributing to the loss of information.
Although this particular aspect may be unique to our simulation, in general, an increased number of subpopulations with different $\eta(Z)$ will improve the inference of $p(Z \mid z)$ at all redshifts.
This is because a given subpopulation $\mathcal{X}$ only gives us access to $p(Z \mid z)$ through the filter of $\eta_\mathcal{X}(Z)$, which may only pick up on a narrow range of metallicities.
In our model, subpopulation $\mathcal{B}$ is only sensitive to metallicities below 0.2 $Z_\odot$ and $\mathcal{C}$ is only sensitive to metallicities below 0.1 $Z_\odot$, with increasing sensitivity at lower metallicities. 
By comparing to $\mathcal{A}$, which is equally sensitive to all metallicities in our example, we directly measure a low-metallicity-weighted SFR below $0.2\,Z_\odot$ and $0.1\,Z_\odot$, respectively. 
The more subpopulations we have, the more quantiles of the metallicity distribution $p(Z \mid z)$ we can directly probe.

It is important to note that even in our three subpopulation case, we can only directly measure the shape of the metallicity distribution below $Z \leq 0.2\,Z_\odot$. At higher metallicities, we are only sensitive to the fraction of star formation at metallicities $Z > 0.2\,Z_\odot$.
The median metallicity falls below this threshold at redshifts $z \approx 10$, and thus is best measured there. 
Otherwise, we infer the shape of the $p(Z \mid z)$ distribution through the specific parameterization we adopted.
The extent of the distribution below the median metallicity informs the high-metallicity end of the distribution as well. 
In particular, recall that we have fixed the parameter that controls the skewness of the metallicity distribution. Leaving this parameter free would widen our constraints on the 95th percentile of the metallicity distribution, because it would no longer be as tightly correlated with our inference of the 5th percentile.  

\section{Discussion}
\label{sec:conclusion}

In just a year of observation, XG observatories will directly measure the BBH merger rate to the few percent level out to $z = 15$. 
Achieving this potential will require overcoming some challenges, including detector technology~\citep{2021arXiv210909882E,2022arXiv221210083S}, data quality and detector calibration~\citep{2024CQGra..41r5001C}, the detection of long-duration, overlapping signals~\citep{2009PhRvD..79f2002R,2023MNRAS.523.1699J,2024PhRvD.109h4015J}, efficient parameter estimation for a large-volume of events~\citep{2021PhRvL.127x1103D,2023ApJ...958..129W}, and managing waveform systematic uncertainties~\citep{2020PhRvR...2b3151P}. 
However, several aspects of data analysis will be simpler than current BBH population studies, due to factors like the absence of selection effects (for mergers with total masses above $10\,M_\odot$) and the large volume of detections, which will enable straightforward, data-driven population inference (e.g., fitting histograms). 

While XG GW data will be incredibly powerful, their astrophysical interpretation relies on robust theoretical models. 
In this work, we focus on the example of inferring the metallicity-specific SFR from XG BBH observations.
As~\citet{2024AnP...53600170C} argued, GW observatories are promising probes of star formation in low-metallicity environments, which is difficult to probe with conventional galaxy observations. 
We build on the work of \citet{2019ApJ...886L...1V} and \citet{2021ApJ...913L...5N}, who showed that the BBH merger rate, as measured by XG observatories, can be used to infer a combination of their progenitor formation rate and delay time distribution.
Following \citet{2023ApJ...957L..31F}, we use theory-informed priors on the delay time distribution in order to better measure the progenitor formation rate.
This progenitor formation rate informs the metallicity-specific SFR, because the formation efficiency of BBH progenitors depends on metallicity.
Critically, the metallicity dependence likely differs between BBH systems with different masses and spins.
We consider a toy model in which multiple BBH subpopulations can be identified via their distinct mass and spin distributions.
Given theory-informed priors on the metallicity-dependent formation efficiency of each subpopulation, we show that comparing their progenitor formation rates allows the overall SFR to be disentangled from its metallicity dependence.

This example underscores the importance of robust theoretical guidance.
We have made the major assumption that the evolutionary histories of BBH systems will be well-understood, so that we can assign informed priors to the delay times distributions $p_\tau$ and metallicity-dependent efficiencies $\eta(Z)$ for the different subpopulations.
By comparing the highly optimistic case that $p_\tau$ and $\eta(Z)$ are known perfectly to a more realistic case where we marginalize over their uncertainties, we showed how a better understanding of BBH evolution improves constraints on the metallicity-specific SFR.  
We expect that these improvements in modeling the formation and evolution of merging BBH systems will stem from an interplay of theory and observation. Theoretical and computational improvements include simulating larger grids of more detailed single and binary stellar evolution models. Observational and data analysis improvements include new multimessenger observations across BBH lifecycles, including binary stars~\citep{2024NewAR..9801694E}, core-collapse supernovae~\citep{2023PASP..135j5002H}, isolated BHs~\citep{2023arXiv230612514L}, BHs in noninteracting binaries~\citep{2024A&A...686L...2G}, accreting BHs~\citep{2021NewAR..9301618M}, and merging BBHs, which will help pin down population synthesis uncertainties. 

In addition to this major assumption, this work made other simplifying assumptions that can be relaxed in future work. 
In particular, we assumed that BBH events can be perfectly grouped into one of three subpopulations.
In reality, BBH masses and spins will have to be simultaneously fit with their merger redshifts.
If distinct subpopulations emerge, BBH systems can be probabilistically assigned to them, which would increase the uncertainty of the measured merger rate of each subpopulation and require a longer observing time to reach a similar level of precision on the merger rates presented in Fig.~\ref{fig:dNdz-histogram}).
Independently of BBH subpopulations, BNS and NSBH systems naturally provide additional populations to compare against the BBH population, and are expected to trace different metallicities.
This was pointed out by~\citet{2024AnP...53600170C}, who proposed comparing the redshift dependence of BBH, BNS and NSBH subpopulations to trace the metallicity-specific SFR in the same spirit as our analysis. 
The toy model presented here uses BBH subpopulations, because BNS and NSBH cannot be detected out to such high redshifts. Nevertheless, NS-containing events will still provide valuable information that should be incorporated, as highlighted by~\citet{2024AnP...53600170C}.   

Alternatively, if there are no clear subpopulations (for example, if one formation channel dominates the BBH merger rate), we still expect trends between the mass and spin distributions of BBH systems and their formation metallicities and delay times (see discussion in \S\ref{sec:intro-subpop}). 
As long as there is some correlation between formation metallicities and BBH properties, we can measure different metallicity-dependent progenitor formation rates and thus disentangle metallicity evolution from the SFR. In fact, smooth correlations may be more powerful at probing the full shape of the metallicity distribution than discrete subpopulations. 
 
While this work focused on inferring the cosmic metallicity evolution, the formation efficiency of BBH mergers may more directly correlate with other conditions of star formation.
For example, if merging BBH systems are dynamically assembled in globular clusters, their progenitor formation rate traces the globular cluster formation rate, which may be correlated with, but distinct from, the low-metallicity SFR. 
If a BBH subpopulation of cluster origin can be identified, one can infer the globular cluster formation rate~\citep{2021MNRAS.506.2362R,2023MNRAS.522.5546F}.
By comparing the cluster formation rate to the SFR (as probed by a BBH subpopulation of field origin), one can then recover the fraction of stars formed in clusters over cosmic history, which is key to understanding galaxy assembly and the specific role of clusters in galactic chemical enrichment~\citep{2023MNRAS.525.4456B}.

Finally, it is important to emphasize the role of multi-messenger observations for understanding cosmic chemical enrichment. While XG GW observatories will probe the low-metallicity SFR to its highest redshifts, they will not be operating in isolation. 
Our understanding of the metallicity-specific SFR will improve significantly in the coming decade, thanks to high-redshift galaxy surveys that are mapping the SFR and mass-metallicity relation, as well as detailed low-redshift studies of Galactic archaeology. 
GW observations provide a highly complementary view into the low-metallicity, high-redshift Universe, as the faintest galaxies often give rise to the loudest GW events.
By the era of XG observations, electromagnetic observations of galaxies will yield better priors on the metallicity-specific SFR that will guide the GW measurements proposed here. 
Furthermore, if one adopts these electromagnetically-inferred SFR relations, the problem addressed here can be inverted, and one can learn about BBH evolutionary histories by inferring the efficiencies and delay time distributions of BBH sources. 
Taken together, the combination of GW and galaxy observations will simultaneously improve our knowledge of cosmic chemistry and GW progenitors.

\ack
I acknowledge support from the Natural Sciences and Engineering Research Council of
Canada (NSERC) under grant RGPIN-2023-05511, the University of Toronto Connaught Fund, and the Alfred P. Sloan Foundation. 
I am also grateful to the Lorentz Center, the scientific organizers and the participants of the workshop ``Gravitational waves: a new ear on
the chemistry of galaxies" (29 April - 3 May 2024,
\url{https://www.lorentzcenter.nl/gravitational-waves-a-new-ear-on-the-chemistry-of-galaxies.html} for many helpful discussions.

\newcommand{\newblock}{}
\bibliographystyle{aasjournal}
\bibliography{references2}

\end{document}